\documentclass[twocolumn,twocolappendix]{aastex63}
\usepackage{amsmath}

\shorttitle{Accretion Flow in M87 is MAD}
\shortauthors{Yuan et al.}

\graphicspath{{./}{figures/}}

\begin{document}
% \linenumbers

\title{The Accretion flow in M87 is really MAD}

%\correspondingauthor{Feng Yuan}
\email{fyuan@shao.ac.cn}

\author{Feng Yuan}
\affiliation{Shanghai Astronomical Observatory, Chinese Academy of Sciences, Shanghai 200030, China}
\affiliation{University of Chinese Academy of Sciences, 19A Yuquan Road, Beijing 100049, China}
%\email{fyuan@shao.ac.cn}
% \item Key Laboratory for Research in Galaxies and Cosmology, Chinese Academy of Sciences, China
\author{Haiyang Wang}
\affiliation{Department of Physics, Fudan University, Shanghai 200433, China}
\author{Hai Yang}
\affiliation{Shanghai Astronomical Observatory, Chinese Academy of Sciences, Shanghai 200030, China}
\affiliation{University of Chinese Academy of Sciences, 19A Yuquan Road, Beijing 100049, China}

%\author{Feng Yuan\altaffilmark{1,2}, Haiyang Wang\altaffilmark{3}, Hai Yang\altaffilmark{1,2}}
%\affil{\altaffilmark{1} Shanghai Astronomical Observatory, Chinese Academy of Sciences, Shanghai 200030, China; fyuan@shao.ac.cn
%\\ \and \altaffilmark{2} University of Chinese Academy of Sciences, 19A Yuquan Road, Beijing 100049, China
%\\\and \altaffilmark{3} Department of Physics, Fudan University, Shanghai 200433, China}

\begin{abstract}
The supermassive black holes in most galaxies in the universe are  powered by hot accretion flows.
Both theoretical analysis and numerical simulations have indicated that, depending on the degree of magnetization, black hole hot accretion flow is divided into two modes, namely SANE (standard and normal evolution) and MAD (magnetically arrested disk).
It has been an important question which mode the hot accretion flows in individual sources should belong to in reality, SANE or MAD. This issue has been investigated in some previous works but they all suffer from various uncertainties. By using the measured rotation measure values  in the prototype low-luminosity active galactic nuclei in {M87} at 2, 5, and 8 GHz along the jet at various distances from the black hole, combined with three dimensional general relativity magnetohydrodynamical numerical simulations of  SANE and MAD, we show in this paper that the predicted rotation measure values by MAD are well consistent with observations, while  the SANE model overestimates the rotation measure by over two orders of magnitude thus is ruled out.  
\end{abstract}


\section{Introduction}
Hot accretion flows around black holes are believed to power low-luminosity AGNs \citep{Ho2008}. They are distinct from the standard thin disks from their much higher temperature, much faster radial velocity, and consequently much lower radiative efficiency \citep{2014ARA&A..52..529Y}. Numerical and theoretical studies have indicated that, in terms of the degree of magnetization of the accretion flow, the hot accretion flows are divided into two modes. One is the standard and normal evolution (SANE), the other is the magnetically arrested disk (MAD) \citep{2003ApJ...592.1042I,2003PASJ...55L..69N,2008ApJ...677..317I,2012MNRAS.423.3083M}. The MAD has attracted much attention in recent years because it is believed to be able to drive  more powerful jets than SANE via the extraction of the black hole spin energy \citep[e.g.,][]{2011MNRAS.418L..79T}. The radiated spectra from SANE and MAD are hard to be discriminated \citep{2019ApJ...887..167X}, but the winds launched from SANE and MAD are significantly different \citep{2021ApJ...914..131Y}, which may result in  significant differences in AGN feedback in galaxy evolution \citep[e.g.,][]{Weinberger2017,2018ApJ...857..121Y}. An important question is then what is the nature of the hot accretion flows in the universe, SANE or MAD?  

Some efforts have been put in answering this question. By investigating two samples of radio-loud active galaxies, good correlations have been found, either between the magnetic flux of the jet and the luminosity \citep{2014Natur.510..126Z}, or between the jet power and the mass accretion rate \citep{2014Natur.515..376G}. These correlations are consistent with the theoretical prediction of MAD theory; therefore it is argued that accretion flow in these sources should belong to MAD in statistical sense. In these works several approximations and assumptions have to be adopted when estimating the physical quantities based on direct observational data, such as bolometric luminosity, black hole mass, jet power, and accretion rate. These uncertainties prevent a firm conclusion to be made.   

When polarization data is available, usually rotation measure, a physical quantity proportional to the integral of the product
of electron density and line of sight component of the
magnetic field, can be used to  constrain the nature of accretion flow \citep{2000ApJ...545..842Q,2014ApJ...783L..33K}. However, when the Faraday rotation and emission occur co-spatially, it is hard to achieve this goal, especially given the turbulent feature of the accretion flow and the complicated general relativity  effects \citep{2020MNRAS.498.5468R}. This is the case of the most recent Even Horizon Telescope (EHT) polarization observations to M87 at 230 GHz, which have obtained the polarized image around the supermassive black hole in M87 on event horizon scales \citep{2021ApJ...910L..12E,2021ApJ...910L..13E}. The low fractional linear polarization is argued to be because of Faraday rotation internal (rather than external) to the emission region, i.e., the accretion flow itself \citep{2021ApJ...910L..13E}. Physically this is because the density and magnetic field within the accretion flow is orders of magnitude higher than that of the external coronal gas. Because of this difficulty, in that work, they instead compare four parameters between the model predictions and those derived from the reconstructed EHT images and ALMA-only (Atacama Large Millimeter/submillimeter Array) measurements.  Intensive survey of model parameters have been performed including black hole spin and the parameter for assigning electron temperature post-hoc in the simulation. In this way, a subset of models have been identified, which can simultaneously explain the most salient features of the observations and produce jet of correct power. It is found that all viable models are MAD \citep{2021ApJ...910L..13E}. But such a result is subject to some model uncertainties. As have been pointed out in that paper, among other things, two uncertainties  are the  electron temperature, which is difficult to precisely determined due to the uncertainties of electron heating from turbulent dissipation and reconnection, and nonthermal electrons in the accretion flow. 

%The magnetic field saturates only within a certain radius. This radius can be regarded as the outer boundary of the MAD. An interesting question is then how large the MAD region is. The 230 GHz radiation mainly come from the innermost region of the accretion flow, $r\la 5 r_g$. 

Therefore, it is useful to continue examining the nature of accretion flow in M87 by invoking additional constraints. We find that the observations of Faraday rotation of jets is very useful for this aim because of the reasons we will state below. Unfortunately, the jets in nearby low-luminosity AGNs are usually very weakly polarized and the Faraday rotation observations are only available to some specific regions in some sources such as Sgr A*, 3C84, and  M87 \citep{2003ApJ...588..331B,2019A&A...622A.196K,2002ApJ...566L...9Z,2014ApJ...783L..33K,2019ApJ...871..257P}. M87 is one of the best targets because of the prototype nature of its hot accretion flow, the presence of the jet, and the abundant observational data. In this work, we make use of the radio polarization observations to the jet in M87. 

By analyzing 8 different  VLBA (Very Long Baseline Array) data sets at 8, 5, and 2 GHz obtained in a range of eight years from 1996 to 2014, \citet{2019ApJ...871..257P} obtained the rotation measure (RM) at various locations in the jet of M87. The locations observed in the jet range from $10^4 r_g$ to $4\times 10^5 r_g$, all inside the Bondi radius, which is $\sim 220 {\rm pc} \sim 8\times 10^5 r_g$ \citep{2018MNRAS.477.3583R}. Here $r_g\equiv GM_{\rm BH}/c^2$ is the gravitational radius of the black hole, $G$ is the gravitational constant, $c$ is the speed of light, and the estimated black hole mass 
$M_{\rm BH}=6.5\times 10^9{\rm~M_\odot}$ \citep{2019ApJ...875L...1E}.
%The latter value is adopted in the present work
The observed RM values are all negative, and systematically decrease with increasing distances from the black hole. No significant temporal variability of RM is detected. The Faraday screen is argued to be located outside of the emission region, i.e., the RM is of external origin, concluded from the its systematic decrease with increasing distance from the black hole \citep{2019ApJ...871..257P}. Theoretical studies have shown the existence of strong wind launched from hot accretion flows \citep{Yuan2012b,Yuan2012a}, and this prediction has been confirmed by growing observational evidences \citep[e.g.,][]{Wang2013,Cheung2016,Shi2021}. Winds have been shown to be responsible for shaping, collimation, and acceleration of jets \citep{Hada2016,Nakamura2018}.  In \citet{2019ApJ...871..257P}, it is further argued that the observed RM is mainly contributed by the wind within the Bondi radius.  The contributions by the Galactic interstellar medium and intergalactic medium in the Virgo cluster are all small and can be neglected, while the contributions from the gas outside of the Bondi radius have been taken into account but it is found to be small \citep{2016ApJ...823...86A,2019ApJ...871..257P}. 

In addition to \citet{2019ApJ...871..257P}, there exist several other polarimetric observations to M87 \citep{2016ApJ...823...86A,Kravchenko2020,2021ApJ...910L..14G}. However, the RM obtained by \citet{Kravchenko2020} is only an upper limit, due to the n$\pi$-ambiguity and differences in time between observations at three observation frequencies. Moreover, since the observations are taken at a position 0.1 mas from the core, very close to the black hole, the accretion flow can simultaneously be the source of synchrotron radiation and the Faraday screen, implying strong depolarization and internal Faraday rotation. \citet{2021ApJ...910L..14G} observed the nucleus of M87 with ALMA. The measured RM exhibits significant changes in magnitude and sign reversals, which is likely caused by the internal origin and the turbulence in the accretion flow. The uncertainties and complexities in the above two works make it difficult to constrain the nature of the accretion flow using the measured RM. On the other hand, the other polarimetric observational results of \citet{2021ApJ...910L..14G} (not RM) have been taken into account in the analysis of \citet{2021ApJ...910L..13E} and shown to be consistent with the MAD nature of the accretion flow in M87. At last, different from the above two works, \citet{2016ApJ...823...86A} observed the M87 jet at locations far beyond the Bondi radius. In this case, the RM is contributed by the interstellar medium rather than the accretion flow, so it cannot be used to probe the nature of the accretion flow.

\section{Model}

To make use of the above observational data, we first simulate both SANE and
MAD accretion flow models of M87. 
Jets are naturally produced in our simulations. We then calculate the predicted RM values at various locations of the jet corresponding to the two
models, and  compare them with the
observational data to see which model is consistent with observations. We will see that the observed external origin of the Faraday screen is consistent with the theoretical prediction, as also been shown by \citet{2010ApJ...725..750B}. This is because the magnetic field of the external medium is comparable to that within the emission region of the jet while the density and length scale of the external medium are much larger. Using these observations not only provides an additional and independent way to investigate  the nature of the accretion flow in M87, it also has important advantage compared to the EHT work. In our case, the calculated value of RM depends only on the simulated quantities of density and magnetic field  which are both robust. The uncertainties due to electron temperature and nonthermal electrons as existed in the EHT work no longer exist. 

Our magnetohydrodynamical numerical  simulations are performed  in a three-dimensional general relativity framework using the ATHENA++ code. Readers are referred to \cite{2021ApJ...914..131Y} for details, here we only briefly describe the key points. A spherical coordinate ($r,\theta,\phi$) is adopted. The inner and outer boundaries of simulation are located at 1.1 $r_g$ and 1200 $r_g$, respectively. We have simulated both SANE and MAD. The observed relatively large jet power in M87 has ruled out all accretion models with a low black hole spin \citep{2019ApJ...875L...5E}, so in the present work we only consider SANE and MAD with a high black hole spin parameter of $a=0.98$. The simulations have lasted for $4\times 10^4 r_g/c$ and $8\times 10^4 r_g/c$ for MAD and SANE, respectively. The ``inflow equilibrium'' has been reached at $\sim (30-40) r_g$. Following \citet{2011ApJ...738...84H}, we have checked the resolution of our simulations by calculating $Q_\theta$ and $Q_\phi$ parameters based on our simulation data, 
\begin{equation}
 Q_\theta = \frac{\lambda_{\rm MRI}}{dx^{\theta}} =\frac{2\pi \lvert v_{\theta,{\rm A}}\rvert}{\Omega dx^{\theta}}, 
\end{equation}
\begin{equation}
 Q_\phi  = \frac{\lambda_{\rm MRI}}{dx^{\phi}} = \frac{2\pi \lvert v_{\phi,{\rm A}}\rvert}{\Omega dx^{\phi}}.
\end{equation}
Here $v_{\phi,A}$ is the $\theta-$ and $\phi-$directed Alfv$\acute{e}$n speed, $\Omega$ is the angular velocity of fluid. The physical meaning of these two parameters is the number of grid in one fastest growing wavelength of MRI in the $\theta$ and $\phi$ directions, respectively. According to \citet{2011ApJ...738...84H}, if $Q_\theta > 10$ and $Q_\phi > 20$, the resolution of the simulation will be high enough to give quantitatively converged results.
We find that the criterions are well satisfied in most region of our simulation domain. A density floor is imposed in our simulations as in all GRMHD simulations. The density floor usually at most affects the density at the jet region. As we will see later, the contribution of the jet to the RM is negligible. So if the density floor adopted in our simulations were too large and a smaller density floor were adopted, the contribution of the jet to the RM would become smaller since a lower density and larger Lorentz factor would be produced thus would not affect our conclusions.

\begin{figure}
	\centering
	    \includegraphics[width=0.6\linewidth]{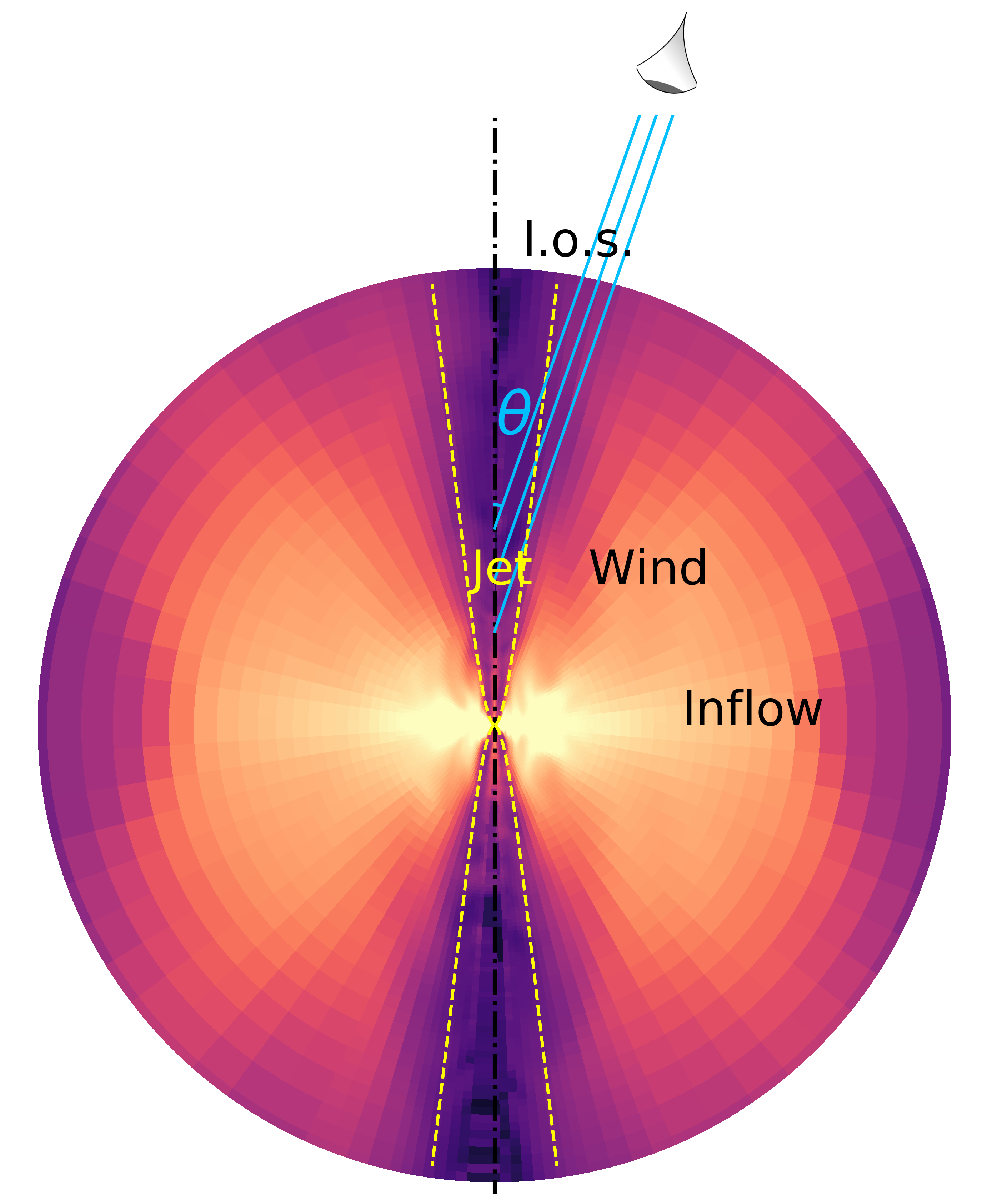}
    \caption{
Schematic figure of the simulated jet-wind-inflow system based on the snapshot data of our GRMHD numerical simulations of MAD98. The color denotes the density of the plasma, the outer boundary of the circle denotes the Bondi radius of M87, while the yellow dashed lines represent the boundary between the jet and wind  (refer to Figure 5 in \citet{2021ApJ...914..131Y}). The viewing angle between the line of sight and the jet axis is $\theta=17^{\circ}$. The integration when calculating the RM is along the solid blue lines.}
\label{fig:jet-wind}
\end{figure}

Figure \ref{fig:jet-wind} shows the overall structure of the simulated system in the $r-\theta$ plane.  Wind and jet have been launched and fill up the whole simulation domain together with the inflow. Since the angle between the line of sight and the jet in M87 is $17^{\circ}$ \citep{2018ApJ...855..128W}, to calculate the value of RM we only need to integrate the plasma with $\theta\la 17^{\circ}$ and this is the region of our interest. The typical velocities of wind and jet in this region is $\sim 0.1-0.2 c$ \citep{2021ApJ...914..131Y}, so the timescale required for the jet and wind to reach the outer boundary  of our simulation domain is $\la 10^4 r_g/c$. This is significantly shorter than the duration of our simulations, therefore the jet and wind launched from within the inflow equilibrium radius up to the outer boundary of our simulation domain have well reached their steady states. 

One may worry that the relatively small inflow equilibrium radius will affect
the properties of jet and wind since the surface density beyond the equilibrium radius may has not settled down to the steady value. However, the extremely long simulation that obtains a large radial range
of inflow equilibrium indicates that there may be no universal surface density within the inflow
equilibrium radius \citep{2020ApJ...891...63W}. This suggests that it is possible that beyond our inflow equilibrium radius the surface density of the accretion flow we obtain may still be realistic thus viable for our evaluation of wind properties. Perhaps more importantly, 
%However, 
the detailed ``virtual particle'' trajectory analysis to outflows (jet and wind) indicates that the outflows located close to the rotation axis are mainly produced at small radii of the accretion flow \citep{2015ApJ...804..101Y}. Given the relatively small viewing angle in M87, we believe that the outflow plasma of our interest mainly come from small radii of the accretion flow and has reached the steady state.   

The observed locations in the jet in \citet{2019ApJ...871..257P}  are much larger than the outer boundary of our simulation domain. Extending the simulation domain to Bondi radius and requiring the gas reaching steady state there is estimated to cost about 5 million CPU hours, which is computationally infeasible. To overcome this difficulty, we have adopted a simplified approach, i.e., extrapolating our simulation data to the Bondi radius.  The detailed approach of extrapolation is described in Appendix A.  Before doing extrapolation to our simulation data, we first azimuthally and time average the data to  eliminate the temporal and spatial fluctuations.  Such a step is even necessary to reflect the behavior of the gas beyond our simulation region, because the degree of fluctuation and turbulence should weaken with increasing distance from the black hole.  This is because the origin of turbulence is the magnetorotational instability which exists mainly in the inflow region thus its effect should become weaker at higher latitude. Such a speculation seems to be confirmed by the different trajectories of virtual particles located at different radii \citep{2015ApJ...804..101Y}. We find that the trajectories of particles located at low latitude are very turbulent while those at high latitude are much less turbulent. This perhaps also explains why no significant temporal variability is detected although these eight sets of polarization observations are taken at different epochs lasting eight years \citep{2019ApJ...871..257P} while significant variability is detected in a one year timescale in the recent polarization observations to the nuclear region of M87 at millimeter bands \citep{2021ApJ...910L..14G}. 

The extrapolation from $\sim 10^3 r_g$ to $\sim 10^6 r_g$ is physically justified. The outer boundary of the hot accretion flow in M87 is located at the Bondi radius \citep{2004ApJ...612..724Y}.  So all the gas properties within the Bondi radius should be controlled by the physics of accretion and outflow, which ensures that our extrapolation is physical. Our extrapolation is also sufficient. All the observed locations in the jet are within the Bondi radius, while as we will show below, our calculations indicate that the RM is mainly contributed by the plasma within the Bondi radius. 

Our simulation data is density scale-free. To calculate RM and compare with the observations, we need to normalize the density and magnetic field of simulations.  We normalize density by requiring  the mass accretion rate calculated based on the simulation data equal to the accretion rate most recently obtained by event horizon telescope collaborations \citep{2019ApJ...875L...5E,2021ApJ...910L..13E}.  
The accretion rate is calculated by
\begin{equation}
\dot{M}=-\int^{2\pi}_0\int^{\pi}_0 \rho u^r \sqrt{-g}d\theta d\phi.
\end{equation}
Here $g\equiv {\rm Det}(g_{\mu\nu})$ is the determinant of the metric, $u^r$ is the contravariant radial 4-velocity. The integration is along the black hole horizon. 
For the MAD model, the mass accretion rate has been determined by combing the single epoch total intensity data, the jet power, and the polarimetric observations \citep{2021ApJ...910L..13E}. The value is $\dot{M}_{\rm BH}=(3-20) \times 10^{-4} M_{\bigodot} {yr} ^{-1}$.  %comparing the resolved polarized structure observed by the EHT, along with the simultaneous unresolved observations with the ALMA (Atacama Large Millimeter/submillimeter Array) to the theoretically expected results\citep{2021ApJ...910L..13E}. The value is in the range of $(3-20)\times 10^{-4}{\rm~M_\odot}{\rm yr}^{-1}$. 
For the SANE model, the mass accretion rate is obtained by using the total intensity data and the jet power \citep{2019ApJ...875L...5E,2021ApJ...910L..13E}. Compared to MAD, unfortunately, the accretion rate of SANE is much less constrained, with  $\dot{M}_{\rm BH}=(5\times 10^{-5}-2.8\times 10^{-3}) M_{\bigodot} {yr} ^{-1}$. The internal energy and the magnetic field strength are also specified once the density is determined. 

%In addition to the constraint from EHTC, there is an additional constraint. The {\it Chandra} observations have measured the density and temperature of the hot gas at the Bondi radius of M87 thus the Bondi accretion rate has been calculated\citep{Russell2015}, which is $\dot{M}_{\rm Bondi} \sim (0.1-0.2) M_{\rm \odot}~{\rm yr}^{-1}$. The radial density profile within the Bondi radius has also been measured, which is consistent with $\rho(r)\propto r^{-p}$ with $p=1.0\pm 0.2$. This corresponds to the radial profile of mass accretion rate of $\dot{M}\propto r^{s}$, with $s=0.3-0.7$\citep{2014ARA&A..52..529Y}. Such a profile  is consistent with MHD numerical simulations\citep{Yuan2012a}, and is due to mass loss of the accretion flow via wind\citep{Yuan2012b}. The deduced accretion rate at the black hole horizon is $\alpha \dot{M}_{\rm Bondi} (1/7.4 \times 10^5)^{s}=(8\times 10^{-7}-3\times 10^{-4})(\alpha/0.1)M_{\rm \odot}~{\rm yr}^{-1}$ with $\alpha$ being the viscous parameter. Combing this constraint with that from EHTC will provide a narrower range of accretion rate. But in the present work we only take into account the EHTC constraint. We will see that even in this case, we already be able to obtain a firm conclusion. 
 
The observed RM is usually smaller than the true value if the beam size of the telescope is moderately larger than the width of the jet \citep{2010ApJ...725..750B}. This is generally the case for the radio core. In this sense, the measured RM there is unreliable. In our case, the observed data is taken from the jet very far away from the radio core, and the beam is smaller than the jet width there \citep{2019ApJ...871..257P}. In this case, the observed RM is very close to the intrinsic value. This is another advantage of using the observational data at large scale to constrain accretion flows, because in this case we don't need to convolve our theoretical calculation  with the beam effect of telescope when compared with observations. 

The RM at different locations of the jet is calculated based on the following formula \citep{2010ApJ...725..750B,2017MNRAS.468.2214M},
\begin{equation}
{\rm RM} =0.812\times 10^6 \int \frac{f(T)n}{g^2}({\rm\bf \hat{k}}-\mathbf{\beta})\cdot {\rm\bf b}dl ~{\rm rad~m^{-2}}.
\end{equation}
%$B_{\parallel}$ is the magnetic field component projected on to the line of sight in unit of Gauss, $dl$ is the line element in unit of $cm$. 
with 
\begin{equation}
g\equiv \Gamma(1-{\rm\bf \beta}\cdot {\rm\bf \hat{k}})
\end{equation}
\begin{equation} f(T)=\frac{1}{\gamma^2}+\frac{\gamma-1}{2\gamma^3}{\rm log}(\gamma),~~~~ {\rm where}~~~  \gamma={\rm max}\left(1,\frac{kT}{m_e c^2}\right). 
\end{equation}
Here $n$ is the electron number density in unit of ${\rm cm}^{-3}$, $T$ is the gas temperature in unit of ${\rm K}$, $\gamma$ is the thermal particle Lorentz factor, $\Gamma\equiv (1-\beta^2)^{-1/2}$ is the bulk gas Lorentz factor, $dl$ is the line element in unit of ${\rm pc}$, $\mathbf{b}$ is the magnetic field in the plasma rest frame in unit of $\mu{\rm G}$ and it is related to the observer frame magnetic field $\mathbf{B}$ by $\mathbf{B}=\gamma(1-\mathbf{\beta} \mathbf{\beta}) \cdot \mathbf{b}$. In our calculations, we have set both $f(T)$ and $\Gamma$ to unity (Appendix B). The integration is along various blue solid lines corresponding to different locations in the jet, as shown in Figure \ref{fig:jet-wind}. 

In the wind, in addition to thermal electrons, in principle non-thermal electrons can also exist, which could be produced by possible magnetic reconnection and turbulence in wind. Eq. (4) is roughly applicable to non-thermal electrons if we interpret $\gamma$ as the mean energy of non-thermal electrons \citep{Huang2011,Dexter2016}. In the present work we only consider thermal electrons, because particle acceleration processes usually only accelerate a very small fraction of thermal electrons into non-thermal distributions. 
%In addition, since non-thermal electrons are accelerated from thermal pool and since the electron temperature in the wind is relatively low as we have argued above, the mean electron energy is expected to be not high. Therefore $\gamma$ and $f(T)$ do not deviate from unity significantly. 

\section{Results}

\begin{figure}
	\centering
	%\begin{subfigure}{0.8\linewidth}
	    \includegraphics[width=1\linewidth]{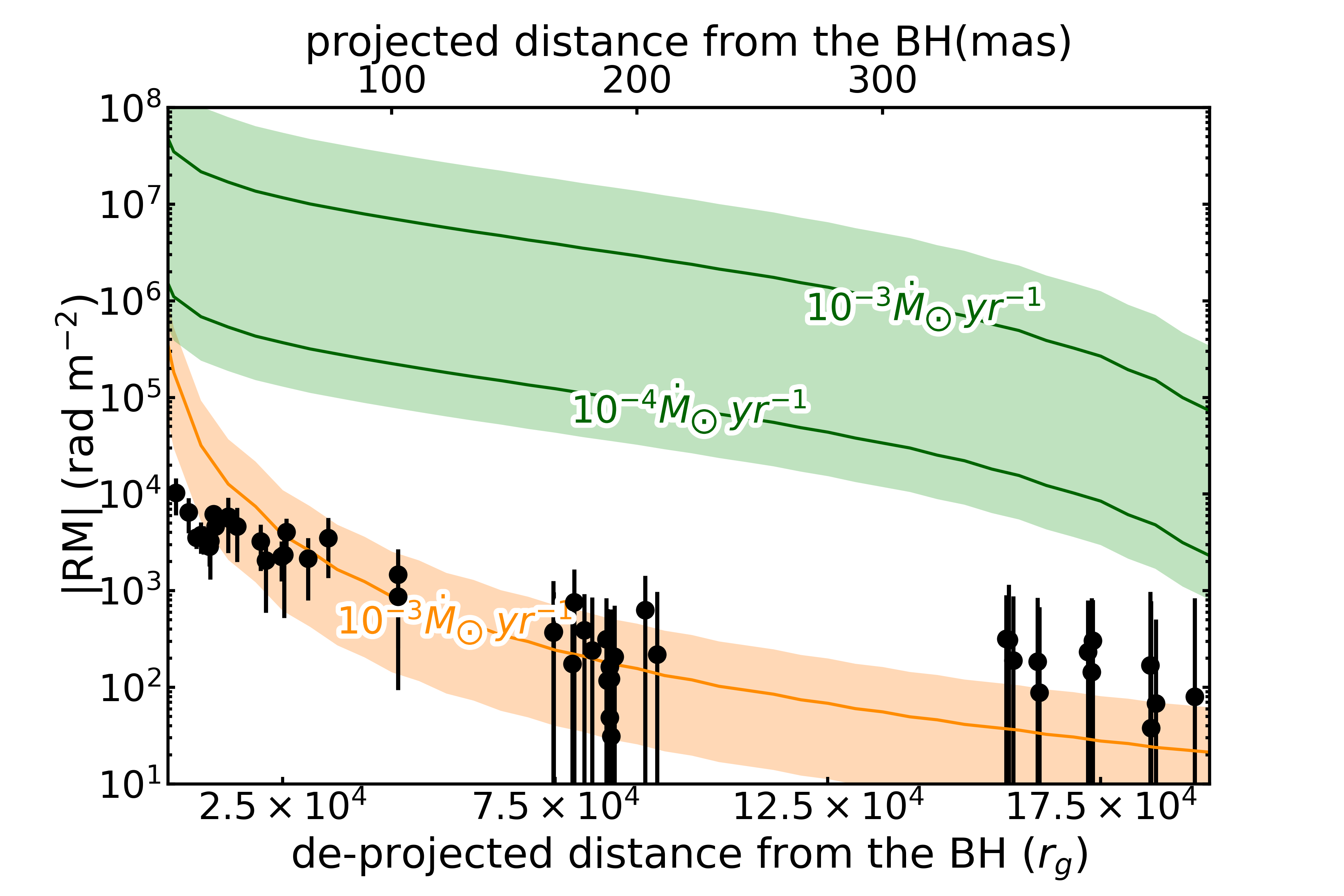}
%	\end{subfigure}\hfil
	\caption{The absolute values of rotation measure as a function of the de-projected distance from the black hole in M87. The observational data is denoted by the black dots with error bars and is taken from Table 2 of \citet{2019ApJ...871..257P}. The orange belt shows the predicted full range of RM by the MAD model which corresponds to the full range of the accretion rate determined by \citet{2021ApJ...910L..13E}. The green belt only shows part of the predicted range of RM by the SANE model which corresponds to the lower range of mass accretion rates determined by \citet{2019ApJ...875L...5E,2021ApJ...910L..13E}, i.e., those corresponding to $a=0.94$ in Table 3 of \citet{2021ApJ...910L..13E}. The values of RM corresponding to other accretion rates are even higher thus not shown here. 
	}
	\label{fig:RM}
\end{figure}

\begin{figure}
	\centering
%	\begin{subfigure}{1.0\linewidth}
	    \includegraphics[width=1\linewidth]{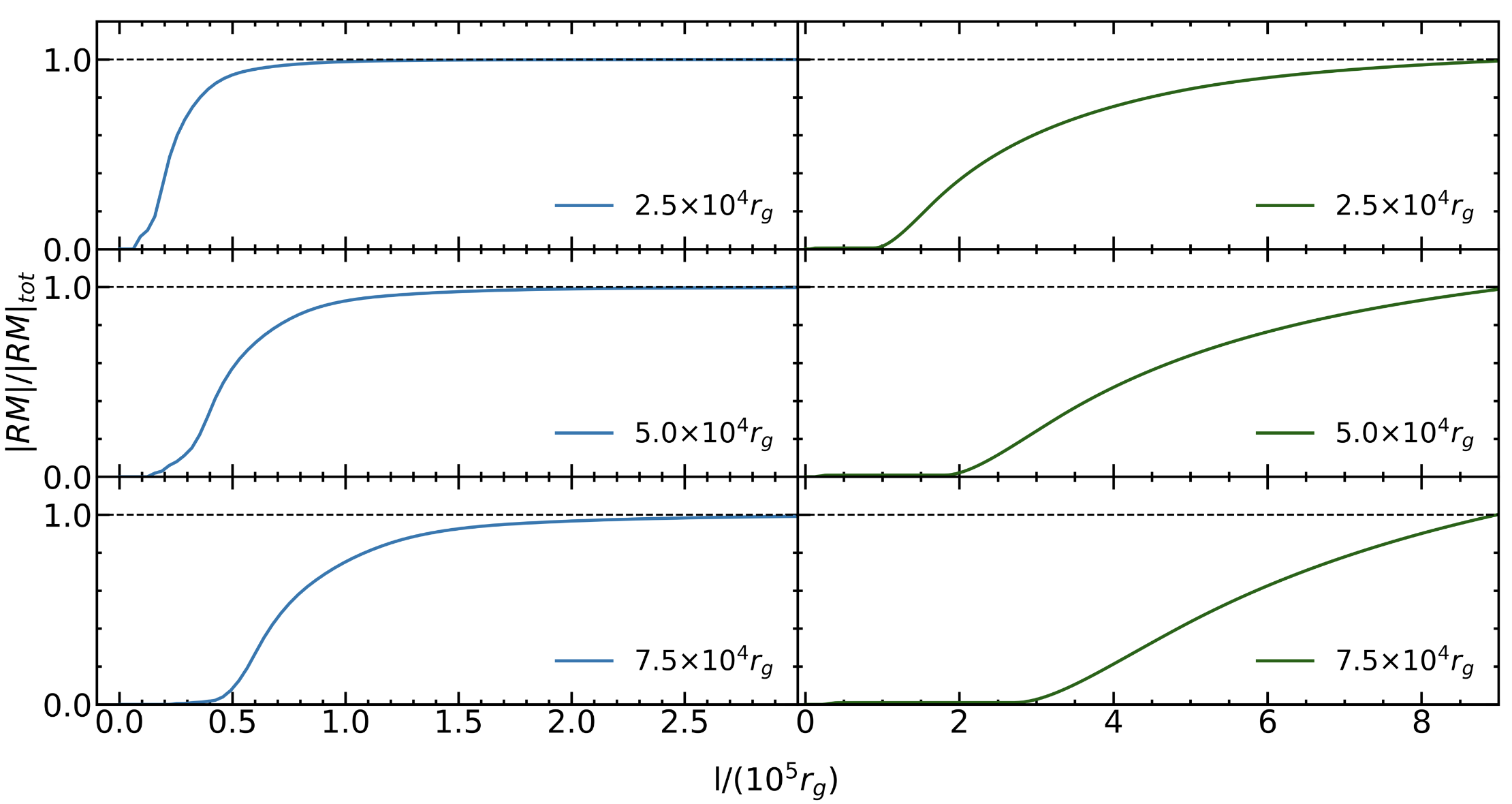}
%	\end{subfigure}\hfil
	\caption{Fractional integrated rotation measure as a function of distance $l$ as measured from the jet  axis along three exemplar lines of sight corresponding to three locations in the jet (as indicated in the figure). The left and right panels are for MAD and SANE, respectively. Combining with the observations of \citet{2019ApJ...871..257P}, these results indicate the external origin of RM. See text for details.}
	\label{fig:fractionRM}
\end{figure}

Our calculations find that the sign of RM at various locations are all negative, consistent with observations\footnote{We note that the sign of RM depends on the orientation of the magnetic field in the outflow region (eq. 4), which in turn depends on the initial orientation of the magnetic field adopted in our simulations. }. Figure \ref{fig:RM} shows the absolute values of RM as a function of de-projected distances from the black hole predicted by SANE98 (by the green belt) and MAD98 (by the orange belt) models and their comparison with observations. Given the very large range of the accretion rate of SANE, we only show the SANE models with the lowest rates. Figure \ref{fig:fractionRM} shows the fractional integrated rotation measure as a function of distance as measured from the jet {\bf axis} along lines of sight. We can see from the figure that more than 80\% of the RM originates from the region beyond $l\sim 2\times 10^4 r_g$ (for the top-left panel) or much larger distances (for the other panels). As comparison, from Figures 2 \& 3 of \cite{2019ApJ...871..257P} we can clearly see that most observed radiation originates from within a distance of (5-10) mas from the jet axis, i.e., within $l=(1.3-2.6) \times 10^3 r_g$. The combination of these results indicates the external origin of RM, consistent with the conclusion in \citet{2019ApJ...871..257P}. From Figure \ref{fig:fractionRM} we can see that in the case of MAD, RM is mainly contributed by the outflow close to the jet axis; while in the case of SANE, RM is dominated by the outflow further away but still within the Bondi radius. It is clear from Figure \ref{fig:RM} that the predicted values of RM by the MAD model are well consistent with  the observed values,  while the prediction by the SANE model is more than two orders of magnitude higher than the observations.

To understand the reason of the difference between SANE and MAD, we have examined the density and magnetic field of MAD and SANE for  the same accretion rate, close to the horizon. We find that depending on the exact location, the density of SANE is slightly higher than MAD while the magnetic field of MAD is a factor of a few stronger than SANE, so their values of RM do not differ too much. The over two orders of magnitude difference of predicted RM between SANE and MAD is mainly because of their different profiles of density and magnetic field. The density and magnetic field in MAD decrease significantly faster with increasing distance than SANE (refer to {Figure \ref{fig:extended3}} in Appendix). Physically, the steepness of the density profile depends on the acceleration of the outflow, which is mainly determined by the Lorentz force because it is the dominant accelerator of outflow \citep{2020ApJ...890...81C}. The magnetization of the outflow in MAD is much stronger than SANE  \citep{2003ApJ...592.1042I,2003PASJ...55L..69N,2008ApJ...677..317I,2011MNRAS.418L..79T,2012MNRAS.423.3083M}, thus the acceleration is significantly larger. The faster weakening of the poloidal magnetic field in MAD compared to SANE can be directly seen from Figure 5 of Yang et al. (2021) by its larger angle between the poloidal field lines and the jet axis.

\section{Discussions and Conclusion}

In the present work, we focus on investigating the degree of magnetization of the accretion flow. Only $a=0.98$ is considered and a parameter survey over black hole spin is not performed. Although the cases of a low black hole spin has been ruled out by the high jet power \citep{2019ApJ...875L...5E,2021ApJ...910L..13E}, the polarimetric analysis in \citet{2021ApJ...910L..13E} indicates that both moderate and high spin values are possible. It is then natural to ask whether  our conclusion changes if a moderate spin value is adopted. Black hole spin will affect the structure and especially the power of the relativistic jet significantly because the jet is powered by extracting the spin energy of the black hole\citep[e.g.,][]{Sasha2012,Narayan2021}. But in our case, the contribution of jet to RM is much smaller than that of wind, as we show in Figure 3. So we expect that the effect of spin via jet is weak. Spin does also affect the properties of wind, which is the dominant contributor of RM. The effects of black hole spin  on properties of wind (e.g., velocity and density) have been investigated in \citet{2021ApJ...914..131Y} by comparing the cases of $a=0$ and $a=0.98$. It is found that the effects of spin are rather weak, much weaker than the effects of degree of magnetization (i.e., SANE or MAD). The reason is that, unlike a jet, the wind is produced at relatively large radii of the accretion flow where the general relativity effects are not very important. The differences between a moderate and high spin values should be weaker compared to the differences between $a=0$ and $a=0.98$. So we expect that spin should not affect our conclusion.

One caveat in our model is the uncertainty of the accretion rates of our SANE and MAD models taken from EHT works \citep{2019ApJ...875L...5E,2021ApJ...910L..13E}. Due to the model uncertainties such as the nonthermal electrons and electron heating, the inferred accretion rates in those two works have some uncertainties. Since roughly we have ${\rm RM} \propto\dot{M}^{1.5}$, to reconcile the discrepancy between the RM-predicted and observed RM, the estimated accretion rate of SANE needs to be lower by a factor of $\sim$ 40. It is highly unlikely that the model uncertainty can produce such a large error. For example, it is estimated that inclusion of non-thermal electrons in the model can only introduce order unity uncertainty \citep{2021ApJ...910L..13E}. 
 
\begin{figure*}
\plottwo{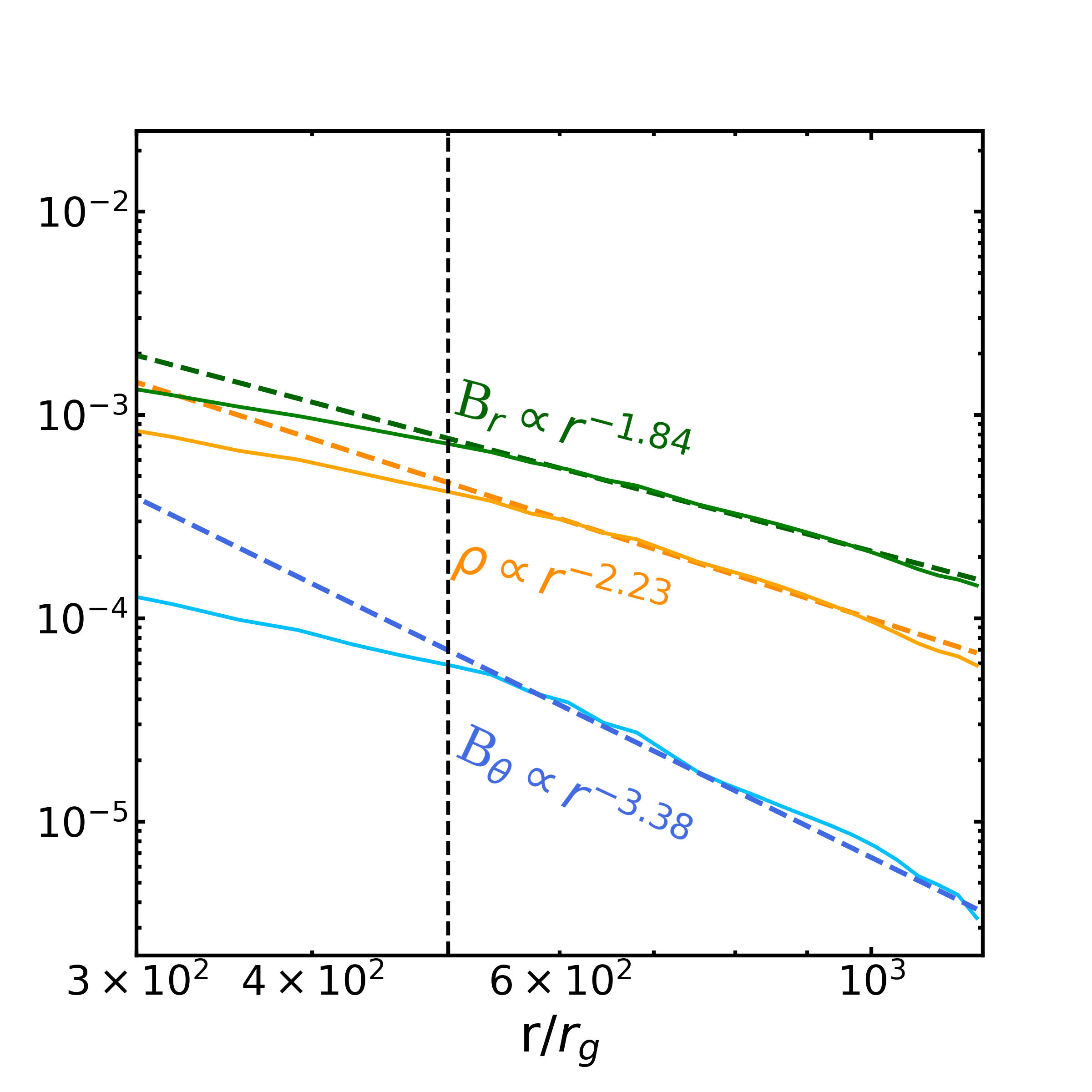}{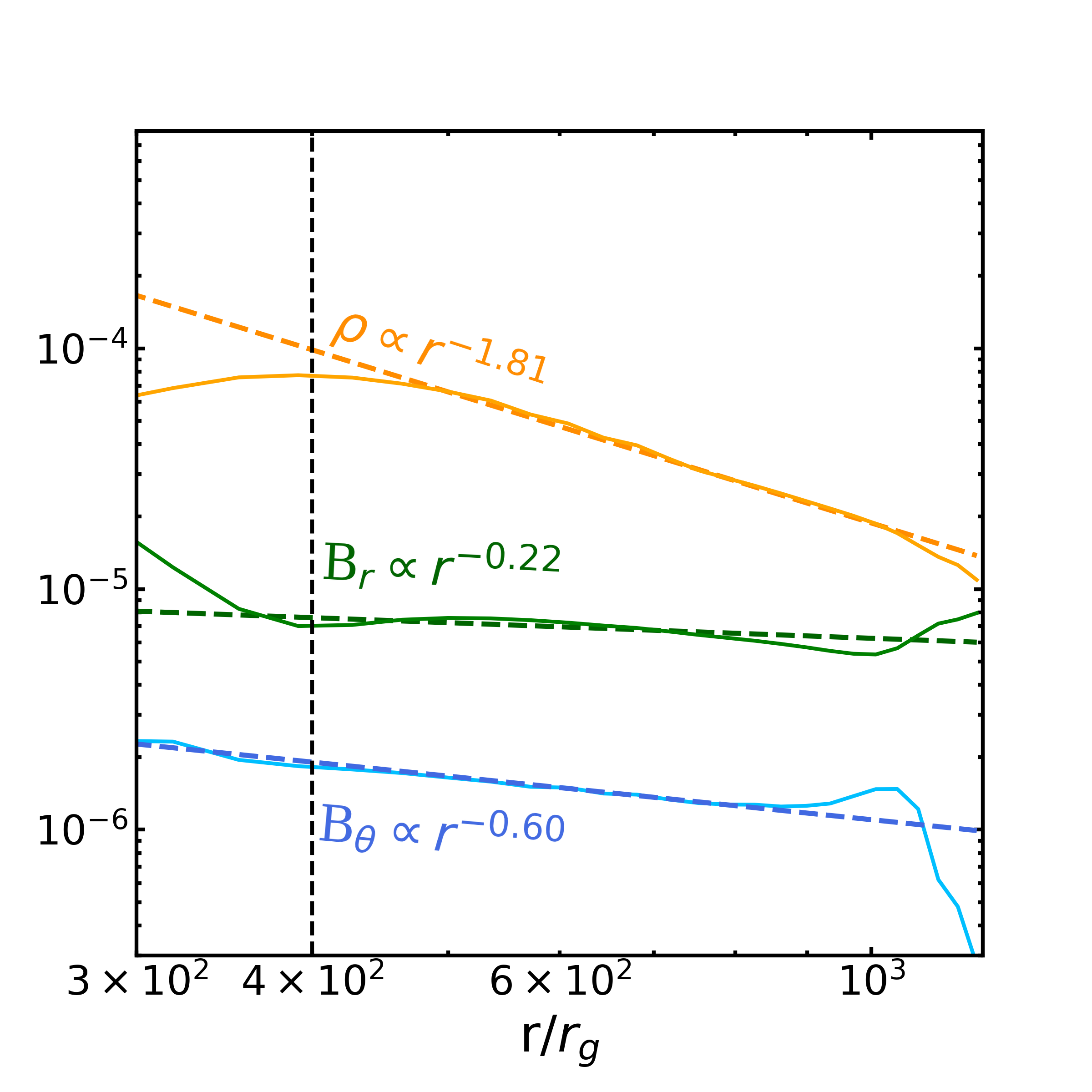}
\caption{The radial profiles of density and $r-$ and $\theta-$ components of magnetic field along $\theta=17^{\circ}$ for MAD (left panel) and SANE (right panel). The solid lines are from our simulations while the dashed ones are their power-law fits beyond the radius denoted by the dashed line.  This radius roughly corresponds to the boundary between the jet and wind (refer to the red lines in Figure 5 of \citet{2021ApJ...914..131Y}). The deviation from the power-law form close to the outer boundary, most clearly displayed in the right panel, is because of the boundary condition effect of the simulation. 
\label{fig:extended3}}
\end{figure*}

The black hole spin adopted in the present work is $a=0.98$, slightly higher than the largest spin value considered in \citet{2019ApJ...875L...5E,2021ApJ...910L..13E}, which is $a=0.94$. The inferred accretion rate is determined by fitting the 230 GHz flux and it is found to decrease with increasing spin, the accretion rate corresponding to $a=0.98$ is thus expected to be smaller than the lower limit of the accretion rates obtained in the EHTC works. However, by comparing the two models with $a=0.5$ and $a=0.94$ (with the same parameter of determining  electron temperature) presented in Table 3 of \citet{2021ApJ...910L..13E}, we can see that their corresponding accretion rates differ only by a factor of $\sim$ 2. It is natural to expect that the difference of the corresponding accretion rates between $a=0.94$ and $a=0.98$ should be significantly smaller. Therefore we expect that the effect of adopting $a=0.98$ should not affect our conclusion. 
 
 The good agreement between the predicted RM by MAD and the observed values provides a strong evidence for the MAD nature of the accretion flow in M87; while the over two orders of magnitude discrepancy between the prediction by SANE and observations rules out the SANE mode. The quantitative consistency between the MAD-predicted and observed RM values in turn provides a strong quantitative support to the predictions of the EHT work on accretion rate and nature of the accretion flow in M87 \citep{2021ApJ...910L..13E}.

\section*{Acknowledgements}
We thank Z.M. Gan and C. White for their helps on numerical simulations using the ATHENA++ code, D.F. Bu, and C. White for their useful discussions, and R. Narayan for his valuable comments.  The constructive comments and suggestions by the referee are acknowledged.
F.Y. and H.Y. are supported in part by  the Natural Science Foundation of China (grants 12133008, 12192220, and 12192223). This work has made use of the High Performance Computing Resource in the Core Facility for Advanced Research Computing at Shanghai Astronomical Observatory. 

\appendix

\section{Extrapolation of simulation data to larger spatial scales}

\begin{figure}
	\centering
	    \includegraphics[width=0.9\linewidth]{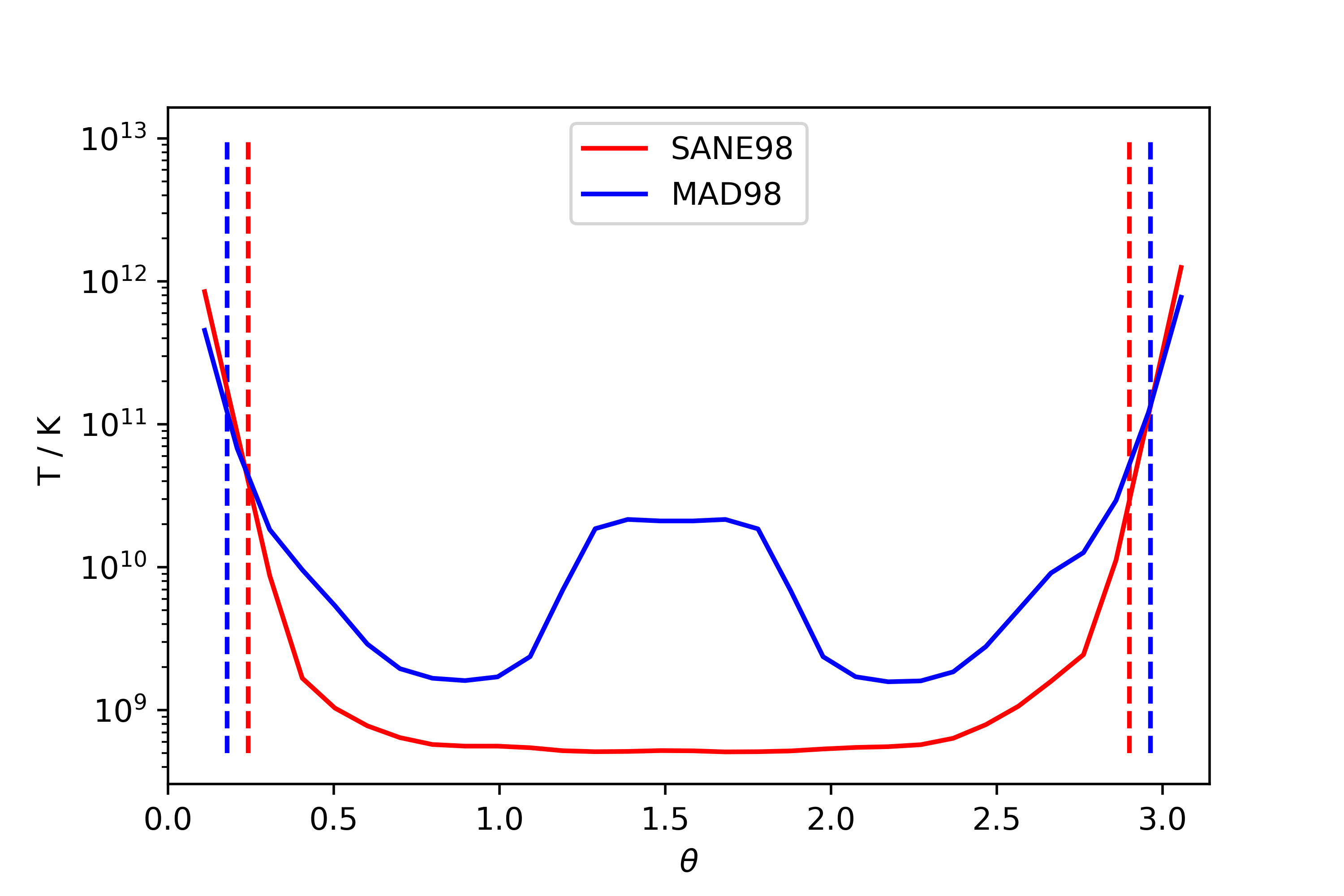}
    \caption{The value of gas temperature as a function of $\theta$ at $1200\,r_g$ obtained in our simulations of SANE98 and MAD98. The vertical dashed line is the boundary of the jet.}
\label{fig:temperature}
\end{figure}

It is found that beyond the fast magnetosonic surface, which is $r_{\rm fms}\sim 100 r_g$, relativistic jets reach the so-called monopole phase  \citep{2009ApJ...699.1789T}. Beyond $r_{\rm fms}$,  the azimuthally and time-averaged physical quantities are very nearly self-similar, with the radial profiles of main physical quantities such as density, internal energy density, components of the velocity and magnetic field following power-law forms \citep{2006MNRAS.368.1561M}. These results are  confirmed by our present simulations. It is found that this is also true in the case of wind \citep{2020ApJ...890...81C}.  So when we do the fitting, we should only adopt the data outside of $r_{\rm fms}$. 

To get the fitting as precise as possible, instead of using a power-law function to describe the radial profile of a certain quantity for the whole $\theta$ region, we have tried to fit  these quantities at various $\theta$ zones individually and do not require their power-law indexes being the same. For some $\theta$ angles, we find that a single power-law function is not enough to present a satisfactory fit but a piecewise power-law (i.e., a broken power-law) is required. This is because, as shown by Figure \ref{fig:jet-wind}, at these $\theta$ angles, our line of sight will pass through both the jet and wind regions in which quantities have different scalings with radius because the properties of jet and wind are controlled by different physics, namely extracting of the black hole spin energy and the rotation energy of the accretion flow, respectively. In this case, we should only choose the power-law function outside of the broken radius when we do the extrapolation of the simulation data from $10^3 r_g$ to $10^6 r_g$. As an example of our fitting, {Figure \ref{fig:extended3}} shows the results for $\theta=17^{\circ}$ for density and the two components of the magnetic field as a function of radius.  The left and right panels are for MAD and SANE respectively. We can see from the figure that both density and magnetic field can be well fitted by  power-law forms, and the slopes of SANE are flatter than those of the MAD. Such a difference is responsible for the differences of the predicted RM between SANE and MAD shown in Figure \ref{fig:RM} and Figure \ref{fig:fractionRM}.

\section{The values of $f(T)$ and $\Gamma$ in eqs. (5) and (6)}
We have set $f(T)$ to unity because the electron temperature in the region of our interest is relatively low thus $\gamma=1$. As shown by Figure \ref{fig:temperature}, at the outer boundary of simulations of both SANE and MAD, the gas temperature is below $3\times 10^{10} {\rm K}$ except within the jet region whose contribution to the RM can be negligible (refer to Figure \ref{fig:fractionRM}). The temperature of wind will decrease with increasing radius, following $T(r)\propto r^{-2(p-1)}$, with $4/3\la p \la 3/2$ being the polytropic index of the wind \citep{Cui2020}. So the temperature of the wind in the region of our interest, i.e., beyond $10^4 r_g$, is no longer relativistic. Similarly, we  simply set $\Gamma=1$ because the velocity of wind is sub-relativistic in that region. 

\bibliography{M87.bib}{}
\bibliographystyle{aasjournal}

\end{document}